\documentclass[sn-nature]{sn-jnl}

\usepackage{graphicx}%
\usepackage{multirow}%
\usepackage{amsmath,amssymb,amsfonts}%
\usepackage{amsthm}%
\usepackage{mathrsfs}%
\usepackage[title]{appendix}%
\usepackage{xcolor}%
\usepackage{textcomp}%
\usepackage{manyfoot}%
\usepackage{booktabs}%
\usepackage{algorithm}%
\usepackage{algorithmicx}%
\usepackage{algpseudocode}%
\usepackage{listings}%

\theoremstyle{thmstyleone}%
\theoremstyle{thmstyletwo}%

\theoremstyle{thmstylethree}%

\begin{document}

\title[DC-operated Josephson junction arrays as a cryogenic on-chip microwave measurement platform]{DC-operated Josephson junction arrays as a cryogenic on-chip microwave measurement platform}


\author*[1]{\fnm{Senne} \sur{Vervoort}}\email{senne.vervoort@kuleuven.be}
\author[1]{\fnm{Lukas} \sur{Nulens}}\email{lukas.nulens@kuleuven.be}
\author[1]{\fnm{Davi A. D.} \sur{Chaves}}\email{daviadchaves@gmail.com}
\author[1]{\fnm{Heleen} \sur{Dausy}}\email{heleen.dausy@gmail.com}
\author[1]{\fnm{Stijn} \sur{Reniers}}\email{stijn.reniers@kuleuven.be}
\author[1]{\fnm{Mohamed} \sur{Abouelela}}\email{M.M.M.Abouelela@tudelft.nl}
\author[2]{\fnm{Ivo P. C.} 
\sur{Cools}}\email{cools@chalmers.se}
\author[3]{\fnm{Alejandro V.} \sur{Silhanek}}\email{asilhanek@uliege.be}
\author[1]{\fnm{Margriet J.} \sur{Van Bael}}\email{margriet.vanbael@kuleuven.be}
\author[4]{\fnm{Bart} \sur{Raes}}\email{bart.raes@imec.be}
\author*[1]{\fnm{Joris} \sur{Van de Vondel}}\email{joris.vandevondel@kuleuven.be}
\affil[1]{\orgdiv{Quantum Solid-State Physics, Department of Physics and Astronomy}, \orgname{KU Leuven}, \orgaddress{\street{Celestijnenlaan 200D}, \city{Leuven}, \postcode{B-3001}, \country{Belgium}}}
\affil[2]{\orgdiv{Department of Microtechnology and Nanoscience},
\orgname{Chalmers University of Technology}, \orgaddress{\city{Gothenburg}, \postcode{SE-412 96}, \country{Sweden}}}	
\affil[3]{\orgdiv{Experimental Physics of Nanostructured Materials, Q-MAT}, \orgname{Universit\'{e} de Li\`{e}ge}, \orgaddress{\street{All\'{e}e du 6 Ao\^{u}t 19}, \city{Sart Tilman}, \postcode{B-4000}, \country{Belgium}}}

\affil[4]{\orgname{IMEC}, \orgaddress{\street{Kapeldreef 75}, \city{Leuven}, \postcode{B-3001}, \country{Belgium}}}


\abstract{Providing radio frequency (RF) signals to circuits working in cryogenic conditions requires bulky and expensive transmission cabling interfacing specialized RF electronics anchored at room temperature. Superconducting Josephson junction arrays (JJAs) can change this paradigm by placing the RF source and detector inside the chip. In this work, we demonstrate that DC-biased JJAs can emit signals in the C-band frequency spectrum and beyond. We fabricate reproducible JJAs comprised of amorphous MoGe or NbTiN superconducting islands and metallic Au weak links. Temperature, magnetic fields, applied currents, and device design are explored to control the operation of the RF sources, while we also identify important features that affect the ideal source behavior. Combined with the proven ability of these JJAs to detect microwave radiation, these sources allow us to propose a fully DC-operated cryogenic on-chip measurement platform that is a viable alternative to the high-frequency circuitry currently required for several quantum applications.
}




\maketitle

\section{Introduction}\label{Sc:Intro}


During the last two decades, cryogenic microwave experimentation for GHz frequencies has evolved tremendously.  The development of the required microwave components is mainly motivated by their strong potential in a variety of quantum applications, including on-chip Ferromagnetic Resonance (FMR), Electron Spin Resonance (ESR), and Nuclear Magnetic Resonance (NMR), as well as research and applications in spin-based and superconducting qubits \cite{Mak15,Pla12,And22,Nie00}. Superconducting materials play an essential role in these experiments, providing the ability to route the microwave signals with low losses and noise. For example, they can be used to fabricate resonators with large quality factors employed in so-called circuit Quantum Electrodynamics (QED) \cite{Bla04}. Currently most of the measurement schemes rely on an interface between these low-temperature superconducting circuits and external commercially available RF components such as arbitrary waveform generators, mixers, oscillators, digitizers, cryogenic amplifiers, circulators, and directional couplers. 

The need for this equipment and the coupling between such a specialized RF circuitry anchored at room temperature to cryostats' cold stages decreases the available space inside the cryostats while hampering both cooling power and system scalability. As such, it is an important task to look for alternative solutions that bring most of these RF components as close as possible to the device under test -- preferably, to the chip itself. The core of many cryogenic electronic components is composed of two weakly coupled superconducting banks, constituting what is known as a Josephson Junction (JJ). Its quantum mechanical nature results in a set of unique electronic functionalities not found in conventional electronics. A JJ can act as an AC current -- DC voltage transducer and can be used as a voltage-controlled oscillator \cite{jos62}. When biased with a fixed DC voltage $V_{DC}$, a resulting AC current flows with frequency  $\nu_J=V_{DC}/\Phi_0$, with $1/\Phi_0 = 2e/h =$ 483.6 MHz/$\mu$V where $\Phi_0$ is the magnetic flux quantum, $e$ the elementary charge, and $h$ Planck's constant. JJ emitters have strong potential as tunable on-chip coherent radiation sources for cryogenic microwave experiments. 
Recently, Peugeot \textit{et al.} successfully realized on-chip spectroscopy using a single Josephson junction (JJ) with an ancillary cavity \cite{Peu24}. Inversely, driven JJs can be used as radiation field detectors as they can respond in a resonant, or phase-locked, manner to the frequency $\nu_{ac}$ of the radiation field they are subjected to \cite{Sha63,Pan20}. This resonant response manifests itself as constant-voltage plateaus, so-called Shapiro steps at integer steps $n$, in the voltage-current ($VI$) characteristics of the junction at voltage values, $V_n=n\Phi_0 \nu_{ac}$. The position of the voltage plateaus $V_n$ depends only on the flux quantum and the driving frequency $\nu_{ac}$.
However, the usefulness of a single JJ for emission and detection applications is limited mainly due to the low response (or radiation power) and small impedance \cite{Wie94,Ben91}. The linewidth of a single junction emission is also limited by the junction resistance. Both these issues can potentially be overcome by using Josephson junction arrays (JJAs) containing many junctions \cite{Til70,Cla73,Jai84,Luk89,Ben91,Wan91,Bi93,Han94,Boo94,Wie94,Kau95,Wel13,Dea17,Cas17}. JJAs provide design flexibility as the number of junctions permits to optimize the array impedance, which allows for better integration \cite{Til70, Jai84,Ben91, Bar99}, while offering robustness against non-uniformities in the JJA \cite{Oct92, Kau95}.
When using JJAs instead of a single JJ, applying a low magnetic field allows for a large in-situ tuning of the junctions critical current \cite{Lob83,Tin83,New00} and, consequently, its radiation power and linewidth -- see Supplemental Information. Another unique property of JJAs is their ability to reach superradiant emission if the several junctions are coherently coupled, or phase-locked, allowing the linewidth to be reduced below the single junction limit \cite{Til70, Jai84, Bar99,Boo94} and the output power to be enhanced significantly \cite{Bar99, Jai84,Wie94, Han94,Boo96}. 

In order to accomplish these promising perspectives, the design and properties of JJAs need to be carefully considered. For instance, it is important to tailor the junctions' critical current and resistance. This is tuned such that an applied current across the junction induces a voltage lying in the interval associated with emission in the desired frequency band. In this work, we explore the unique properties of 2D planar Josephson junction arrays for on-chip generation and detection of microwave signals for quantum information processing. By fabricating JJAs with different materials and designs, we demonstrate Josephson radiation in the complete C-band and beyond. The frequency tunability is achieved by biasing the arrays with DC currents that induce an average voltage over the junctions. We discuss the impact of different fabrication choices on the superconducting properties of the arrays and reveal how a crucial interplay between the emitted radiation and the circuitry transmitting this signal affects how it is measured. Finally, we discuss important factors related to the radiation source performance and demonstrate that the same arrays can be used as detectors for radiation in the GHz range. This allows us to propose a novel GHz spectroscopy device operated completely by DC sources, eliminating the need for RF circuitry altogether.

\section{Results}\label{Sc:Results}

The C-band of microwave frequencies, spanning from 4 to 8 GHz, is a sweet spot for circuit QED and quantum computing applications. From a practical perspective, RF electronics is commonly optimized to work in this frequency range. From a fundamental viewpoint, the C-band lies well below the superconducting gap frequency avoiding quasiparticle poisoning, and is sufficiently high in frequency to ensure that effects of thermal energy on the resonator state remain negligible at millikelvin temperatures. As a proof of principle, the 4--8 GHz range will be the targeted frequency window for this study. As the emission frequency is linked to the voltage over the junction via the Josephson relations, the envisioned JJA design should have the required fabrication flexibility to tailor the weak link critical currents ($I_c^{wl}$), associated with the onset of the dissipative regime in the array, and the dissipative state average single junction resistance ($R_j$) \cite{Wie94,Lik79}. This enables controlled access to the voltage state in the range $V_j = V_{DC}/(N_x-1) \approx 8.27 - 16.54~\mu$V corresponding to the Josephson radiation frequency $\nu_j = \frac{V_j}{\Phi_0}$ within 4--8 GHz, where $V_j$ stands for the average voltage across an individual JJ along the bias direction $x$, and $N_x$ is the number of islands along the $x$-direction. Moreover,  the fabrication process should be highly reproducible, enabling the fabrication of arrays of nearly identical junctions \cite{Jai84,Lik79}. 
 
 \begin{figure}[h]
    \centering
    \includegraphics[width=1\linewidth]{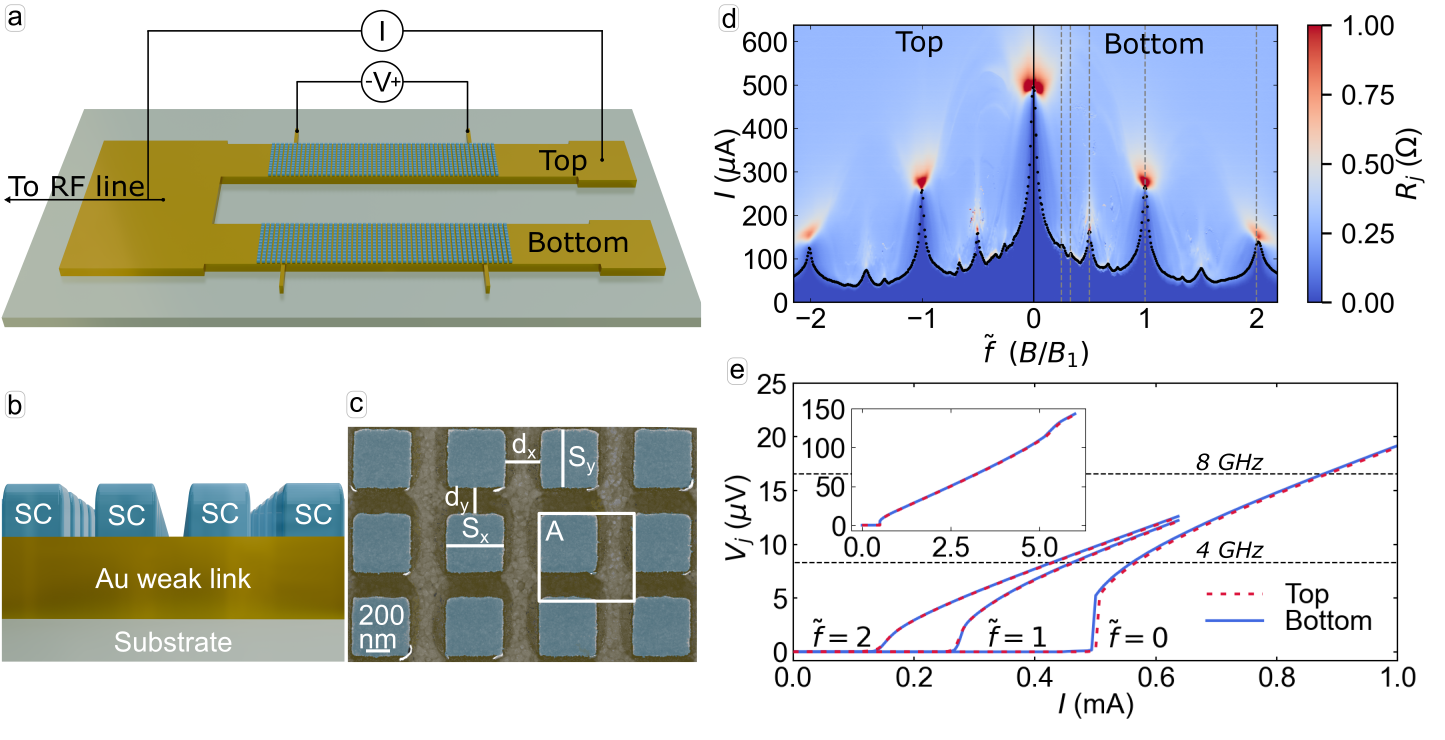}
    \caption{\textbf{Josephson junction array device with magnetic field-tunable $VI$ curves.} \textbf{a} Schematic of the devices, consisting of two metallic transport bridges each containing a superconducting JJA, labeled top and bottom. The arrays can be individually biased with a DC current and share an RF pickup contact. Separate contacts are present to measure the voltage over the junctions. \textbf{b} Schematic overview of the used stack. The superconductor is indicated in blue and the gold weak links in yellow. \textbf{c} False-colored SEM image of a representative array (MoGe-top). The design parameters $S_x$, $S_y$, $d_x$, $d_y$ and the resulting unit cell area $A$ are indicated in the figure. \textbf{d} Average junction dynamical resistance, $R_j$ as a function of applied DC current, $I$, and frustration, $\tilde{f}$, for the top (MoGe-top) and bottom (MoGe-bot) array at 300 mK. The weak link critical current, $I_c^{wl}(\tilde{f})$ is shown as black dots. Grey dashed vertical lines sequentially indicate $\tilde{f}$ = 1/4, 1/3, 1/2, 1, and 2. \textbf{e} Average junction voltage, $V_j$, as a function of applied DC current $I$ for the top (red dashed) and bottom (blue) arrays, at $\tilde{f}$ = 0, 1, and 2. Horizontal black dashed lines indicate junction voltages according to the relation $V=\Phi_0 \nu_{ac}$ with $\nu_{ac} = 4$  and 8 GHz (the C-band). The inset shows the $VI$ curve for $\tilde{f} = 0$ up to 6.5 mA.}
    \label{fig:1}
\end{figure}

To achieve the above requirements, we fabricate devices as schematically presented in Fig.~\ref{fig:1}a -- for details on the fabrication steps, see Section \ref{Sc:Methods}. SEM images of this device can be found in SI Fig. 9. They consist of two normal metal Ti/Au transport bridges, each decorated with an array of regularly spaced square superconducting islands, either of Au-capped Mo$_{78}$Ge$_{22}$ or of NbTiN, thus defining superconducting-normal metal-superconducting (SNS) planar junctions in the long diffusive regime as shown in Figs. \ref{fig:1}b and c \cite{Lik79,Pan20}. The islands have nominal sizes $S_x$ = $S_y$ = 500 $\pm$ 25 nm and are separated by $d_x$ = 100, 200, 300, or 400 nm and $d_y$ = 200 nm, as depicted in the SEM image in Fig. \ref{fig:1}c. Each array contains $N_x$ and $N_y$ islands in the $x$ and $y$ direction respectively, thus presenting $(N_x-1)\times(N_y - 1)$ junctions. Including two bridges in each device allows us to test the reproducibility of our fabrication process, as well as the uniformity of the superconducting behavior, by separately biasing each array. In total, we have investigated 10 JJAs and, in the remaining of this work, we will report on selected representative devices, as listed in Section \ref{Sc:Methods}. The reported behaviors were consistently reproducible on all studied samples.

Using SNS junctions instead of tunnel junctions brings some important advantages to the design. Firstly, SNS junctions demonstrated a broad dissipative state far below $T_c$, defined at the onset of the transition to the non-dissipative state as the sample is cooled down \cite{Ele13,Pan20}. Secondly, $I_c^{wl}$ in SNS junctions depends decisively on the ratio between the interisland spacing and the normal metal coherence length ($\xi_N$) \cite{Lik79}. As the latter length scale ($\xi_N(T_c) \approx 70$ nm) is within the range of state-of-the art fabrication processes, it provides a design knob to tune the device response. Finally, SNS junctions eliminate the need for an oxide tunnel junction, a source of two level system losses in quantum computing schemes \cite{VaDa23,Mül19,Phi87}. Thus, designing devices with such JJAs allows tuning the emitted power depending on the application, by simply adjusting the total amount of junctions and their arrangement ($N_x$, $N_y$, $S_x$, and $S_y$) \cite{Luk89,Ben91}. 

To test reproducibility and sample uniformity, we probe the magnetic field ($B$) dependence of the $VI$ characteristics of separate arrays within a single device. Two mechanisms contribute to the $R_j(IB)$ response, shown in figure \ref{fig:1}d. 
Firstly, the critical current of individual junctions exhibits a Fraunhofer-like pattern as a function of $B$, with periodicity related to the junction's area \cite{Tin96}. Secondly, there is an interplay between the superconducting vortices and the two-dimensional array. When the vortex lattice is commensurate with the array, vortices are distributed forming a periodic stable configuration, resulting in an increased $I_c^{wl}$ \cite{Tin83,New00,Rzc90}. The magnetic fields at which this happens are called matching fields, defined as $B_k  = k\frac{\Phi_0}{A}$, with $\Phi_0$ the magnetic flux quantum and $A$ the unit cell area of the JJA (see Fig. \ref{fig:1}c) and $k$ a rational number. The value of $k$ represents the average number of flux quanta contained per unit cell. It is convenient to express the magnetic field acting on the array in reduced units, known as frustration, $\tilde{f} = \frac{BA}{\Phi_0} = \frac{B}{B_1}$. 

Figure \ref{fig:1}d simultaneously presents the average single junction resistance $R_j = dV_j/dI$ of two identically designed MoGe arrays on the same device (MoGe-top and MoGe-bot) at 300 mK. The behavior is investigated as a function of bias current in magnetic fields between $-9$ and $9$ mT. The results are obtained by measuring $VI$ curves at different fields with 30 $\mu$T steps. For negative $\tilde{f}$ values, we present results for the top array, while those for the bottom array are shown for positive $\tilde{f}$ (see Fig. \ref{fig:1}a for reference). If the arrays have equivalent properties, the matching peaks should be mirrored around $\tilde{f} = 0$, as evidenced in the figure. The black dots mark $I_c^{wl}$, determined as the current value at which $V_j = 1~\mu$V. Two features are immediately recognized. First, results from both arrays are nearly indistinguishable, with $I_c^{wl}(0)$ equal to 495 $\mu$A and 493 $\mu$A for the top and bottom bridge, respectively. Second, clear matching features are observed at both integer and fractional (i.e. 1/2, 1/3, ...) frustrations, a signature of the uniformity and negligible disorder in the nanofabricated arrays. The presence of pronounced matching peaks in the $I_c^{wl}(\tilde{f})$ curves in Fig. \ref{fig:1}d that are symmetrical with respect to $\tilde{f} = 0$ is thus an indication of uniformity in fabrication and superconducting properties across the array.

For $I > I_c^{wl}$, the arrays enter the dissipative state and the operational emission conditions \cite{Luk89,Wie94}. In Fig. \ref{fig:1}e, we investigate individual $VI$ curves at different matching fields for the top (dashed red lines) and bottom bridges (solid blue lines). For $\tilde{f}$ = 0, 1, and 2, the observed $V_j$ falls into the voltage range associated with Josephson radiation in the C-band, represented by the two horizontal dashed lines (with $R_j = 0.33$ $\Omega$ in the linear section). It is important to stress that this voltage match with the C-band frequency range is the result of careful fabrication and design choices, tuning $I_c^{wl}$ and $R_j$ for a given temperature. Nevertheless, at $\tilde{f} = 0$, the voltage-current characteristic is slightly non-linear close to the values corresponding to emission at 4 GHz. It is known that non-linearity in the $VI$ behavior will lead to undesired higher harmonics in the radiated frequency spectrum \cite{Luk89}. Another problem we identify (see Supplemental Information) is that the radiation is highly sensitive to voltage differences between different junctions in the non-linear part of the $VI$ curve. Therefore, the radiation produced in this regime has a larger linewidth and decreased intensity at the desired frequency. To mitigate this problem, one can benefit from the inherent magnetic field commensurability effects at $\tilde{f}=1$ and 2 to access a more linear section of the $VI$ curve within the C-band, while maintaining vortex stability and avoiding detrimental effects of flux flow. The inset of Fig. \ref{fig:1}e shows $VI$ curves for the top and bottom arrays spanning the full voltage state. The dissipative state sets in around 0.5 mA, when the junction’s critical current is exceeded. A second transition just above 5 mA, i.e. much higher than the required current to operate the device in the desired frequency range, indicates the critical current of the superconducting islands and a transition to the normal state, which is found to be almost identical for the MoGe islands in the top and bottom JJAs [5.15 mA (top) and 5.13 mA (bottom)]. This current and voltage range also sets the upper frequency limit for the emission of these particular MoGe devices around 55 GHz ($V_{DC} = 114~\mu$V), indicating that it is possible to use JJAs as high-frequency radiation sources for different applications.

\begin{figure}[h]
    \centering
    \includegraphics[width=1\linewidth]{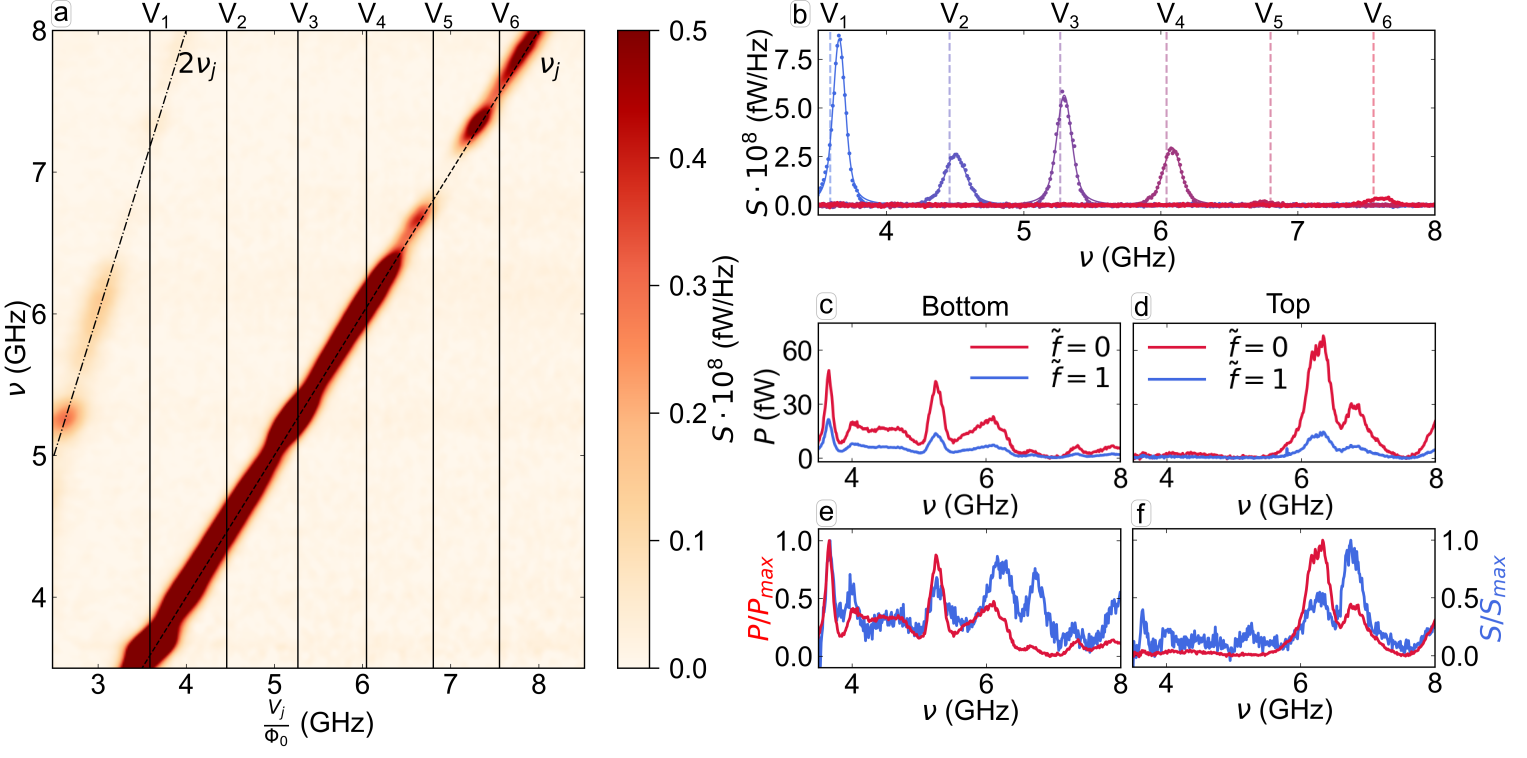}
    \caption{\textbf{DC voltage-tunable Josephson emission.} \textbf{a} Power spectral density $S$ as a function of frequency $\nu$ and junction voltage $V_j$, with $V_j$ converted to units of Hz using the Josephson equation. Data is for MoGe-bot at 300 mK and $\tilde{f} = 1$. \textbf{b} Selected spectra of the radiation map in panel a (solid symbols) and fits (solid lines). Vertical dashed lines indicate the expected Josephson frequency. \textbf{c} Integrated power $P$ over bias voltage for MoGe-bot (\textbf{d} top) as a function of frequency $\nu$ at different values of frustration. \textbf{e} The power integrated over voltage (red) and the measured white noise power spectral density (blue), both normalized to their maximum value for MoGe-bot (\textbf{f} top). The white noise is an average of 25 measurements with a biasing current of 6000-6500 $\mu$A at a frustration of $\tilde{f} = 0$.}
    \label{fig:2}
\end{figure}

After characterizing the arrays' response to a DC current bias, we tested their ability to generate Josephson radiation in the 4-8 GHz range using a signal analyzer, as described in Section \ref{Sc:Methods}. 
The power spectral density $S(\nu)$ is measured as a function of $V_j/\Phi_0$, where $\nu$ is the probed frequency. Figure \ref{fig:2}a shows results for MoGe-bot operated at 300 mK and $\tilde{f} = 1$. The regions of higher $S$ (shown in darker brown) follow the expected Josephson frequency's first harmonic, highlighted by the black dashed line labeled $\nu_j$. A less intense second harmonic signal can also be observed at $2\nu_j$. Simply by varying the DC voltage across the array, we are able to produce a measurable and tunable radiation in the GHz range transmitted through the normal metal transport bridge.

Figure \ref{fig:2}b further explores the power spectral density by showing six emission spectra at equally spaced voltage intervals indicated in panel a, by vertical lines and named $V_1$ until $V_6$. In panel b, solid lines are Voigt fits of each observed radiation peak, while vertical dashed lines represent the expected Josephson frequency for each spectrum. The spectrum at $V_1$ is chosen to represent the value at which the measured radiation power is maximum, $P = P_{max}$. $P$ is obtained by numerically integrating $S(\nu)$ over the full frequency range. In this case, the power is 11.9 fW and the full width at half maximum (FWHM) is 106.5 $\pm$ 0.1 MHz. This leads to a maximal power efficiency, defined as the ratio between the input and output powers, around $10^{-7}$, which is consistent with the efficiency observed for tunable radiation originating from vortex movement in superconducting superlattices \cite{Dob18}.  Importantly, the observed radiation from the source is not coherent as the linewidth of the detected signal is much larger than that theoretically expected for a single junction, which is around $\Delta \nu \approx 4$ MHz for these devices \cite{Ste69,Luk89,Wie94}.

Figure \ref{fig:2}c shows $P(\nu)$ curves, obtained by integrating the signal over all voltages, for array MoGe-bot at $\tilde{f}$ = 0 and 1, evidencing a decrease in the magnitude of $P$ as $\tilde{f}$ is increased -- similar to the behavior of $I_c^{wl}$ shown in Fig. \ref{fig:1}e and in accordance with results obtained from the current driven resistively shunted junction (RSJ) model for a single junction \cite{Pan20,Rae20} (see Supplemental Information).
In addition to this, the RSJ model predicts (i) that the radiated power increases smoothly and monotonically as the voltage across a Josephson junction increases; and (ii) that the power of the second harmonic radiation is proportional to the first harmonic component. In contrast to these predictions, Fig. \ref{fig:2}c shows a highly non-monotonic behavior for $P(\nu)$. Moreover, inspecting $S$ along the first harmonic emission in Fig. \ref{fig:2}a, we observe a maximum at $V_1$. If the maximum detected power was related directly to a stronger emission at this particular voltage, the second harmonic should also be enhanced. However, following the vertical $V_1$ line in panel a, we notice that the second harmonic emission is not enhanced. Instead, a clear power enhancement is observed for the second harmonic at $\nu$ = 5.2 GHz. At the same $\nu$, we observe an enhanced power for the first harmonic along $V_3$ -- see also Fig. \ref{fig:2}b. This indicates that the modulations are related to the frequency rather than to the junction voltage. 

This is further supported by comparing the results from Fig. \ref{fig:2}c with those from Fig. \ref{fig:2}d, which are obtained for  MoGe-top. As shown in Fig. \ref{fig:1}d, the two arrays have analogous superconducting responses, suggesting that the different $P(\nu)$ curves observed are not related to the profile of the Josephson radiation emitted by the arrays, but result from a frequency-dependent transfer function related to the transport bridge, wirebonds, and the sample holder that connect the emitter (JJA) and the receiver (output line). 
This hypothesis is reinforced by the results in Figs. \ref{fig:2}e and f.
The red curves represent the same data as in panels c and d, but here they are plotted normalized to $P_{\text{max}}$. In blue, we present the normalized power spectral density, $S/S_{\text{max}}$, observed when the Josephson radiation is replaced by a white noise radiation source (see Methods). As $S/S_{\text{max}}$ presents features closely related to $P/P_{\text{max}}$ for both arrays, the modulation in the detected power are independent of the source and thus not related to the emission of the JJAs. The differences in the radiation profile depending on the position of the array in the device (top or bottom in Fig. \ref{fig:1}a) were consistently observed for all measured samples.

\begin{figure}[h]
    \centering
    \includegraphics[width=1\linewidth]{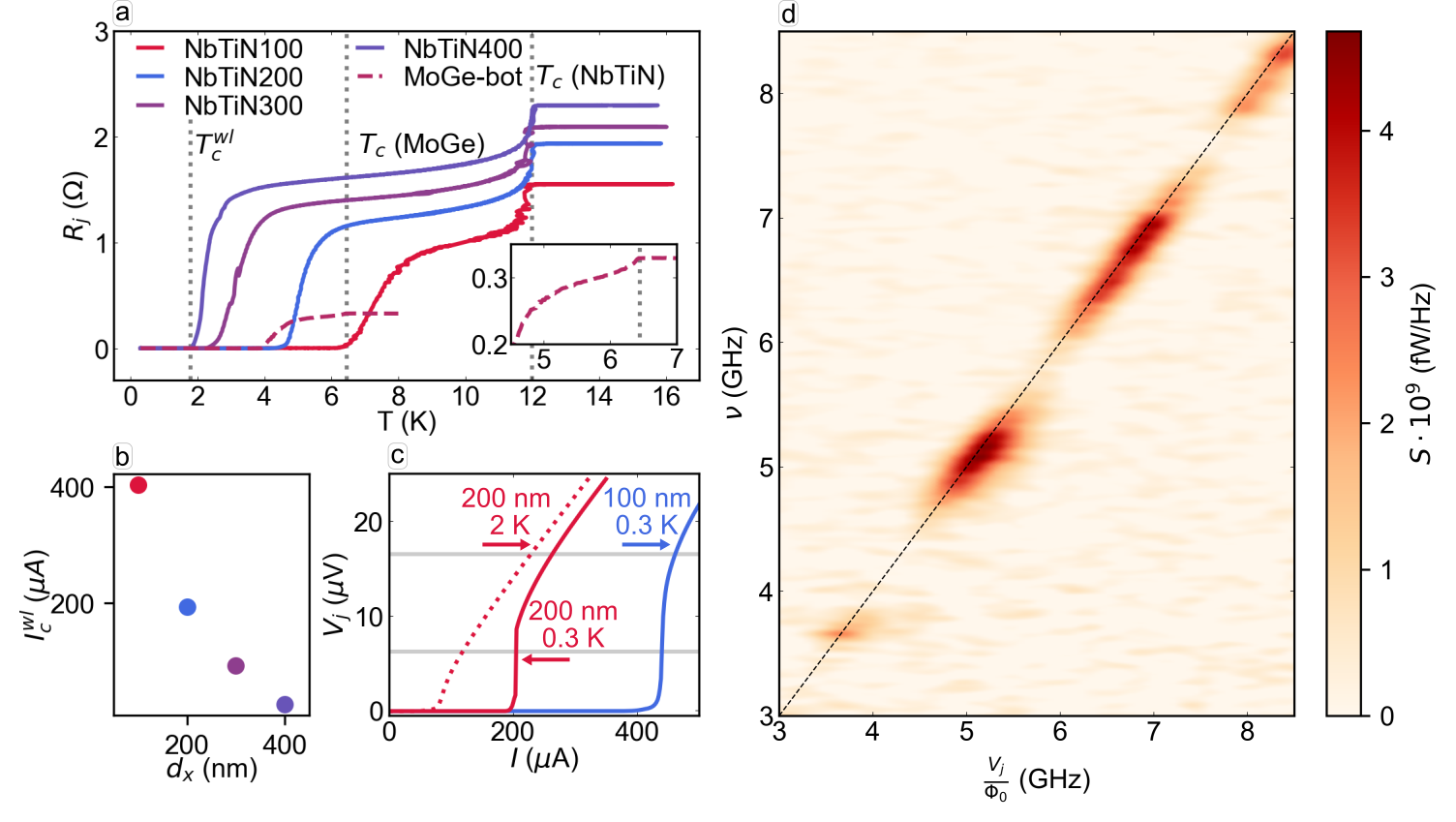}
    \caption{\textbf{Effect of $d_x$ and temperature on DC voltage-tunable Josephson emission in NbTiN JJAs.}  \textbf{a} Junction resistance as function of temperature for NbTiN100 - 400 and MoGe-bot. The $T_c^{wl}$ is indicated by a grey dashed line for the array with $d_x = 400$ nm, and the superconducting $T_c$ is the same for all devices $T_c = 12$ K. The inset shows the dependency of $I_c^{wl}$ on $d_x$. \textbf{b} \textit{VI} curves for NbTiN100 and NbTiN200 at different temperatures. \textbf{c} Power spectral density $S$ as a function of frequency $\nu$ and junction voltage $V_j$, converted to units of Hz using the Josephson equation. Data is for  NbTiN200 at 2 K.}
    \label{fig:3}
\end{figure}

The previous results establish that MoGe/Au SNS JJAs can be operated as DC voltage-tunable GHz radiation sources. As shown in Fig. \ref{fig:3}a and in the inset, the amorphous MoGe JJAs have a $T_c$ = 6.5 K, defined at the onset of the transition to the dissipative state as the sample is cooled down. Nevertheless, MoGe has a few important drawbacks as a material in nanodevice applications and superconducting circuitry. Particularly, the superconducting properties (like the critical temperature) are strongly dependent on the exact stoichiometry of the MoGe compound. To control the stoichiometry, a pulsed laser deposition is needed during fabrication, which limits the uniformity of the thin films and compromises the use of lithographic steps. To tackle these issues, boost the critical temperature of the chosen superconductor, and test the robustness of the radiation mechanism in a material that is typically employed for quantum technology, we fabricated new devices in NbTiN \cite{Schuck2013,Miki2013,Van2015,Schuck2016,Chaudhuri2017,Hazard2019}. This material has a bulk $T_c$ of 17.3 K \cite{Fel16} and around 14 K for films with thickness of 20 -- 40 nm \cite{Zha15}. The NbTiN islands are embedded on similar Au transport bridges as the MoGe devices using a top-down approach as described in Section \ref{Sc:Methods}. While all devices have the same number of junctions, the horizontal interisland spacing $d_x$ is varied from 100 nm to 400 nm.
Figure \ref{fig:3}a shows the $R_j(T)$ characteristics of these samples, where the bias current is far below the critical current. The curves present two transition temperatures, as expected for normal metallic films decorated with coupled superconducting islands \cite{Ele12,Ele13}. At $T_c$ = 12 K, for all devices, a normal state-superconducting transition indicates that the NbTiN islands retain superconductivity despite the presence of the underlying and capping Au layers. At lower temperatures, labeled as $T_c^{wl}$, the devices enter a non-dissipative state as the superconducting islands are effectively coupled by the underlying Au bridge, such that $T_c^{wl}$ sets the upper limit at which the devices can emit Josephson radiation. We observe that $T_c^{wl}$ decreases from 6.4 K to 1.8 K as $d_x$ increases from 100 nm to 400 nm, highlighting the effect of the interisland spacing on the coupling between the islands. Fig. \ref{fig:3}b further demonstrates this effect, as the $I_c^{wl}$ decreases with increasing $d_x$.
Another important ingredient influencing the coupling is the normal metal coherence length ($\xi_N$) \cite{Ele12}. As $\xi_N$ is directly related with the electronic mean free path via $\xi_N \approx0.58 \sqrt{\frac{\hbar v_f}{2\pi k_BT}l}$ \cite{Lik79}, the resistivity of the Au bridge affects the interisland coupling and $T_c^{wl}$. The data in Fig. \ref{fig:3}a corroborate that trend as we observe a large difference between $T_c$ and $T_c^{wl}$ for the NbTiN samples, for which the Au bridge has a resistivity around 53 $\Omega$nm. For comparison, the Au bridge in the MoGe samples has a resistivity of 16 $\Omega$nm, 3.3 times lower than what we obtained for the NbTiN samples. Thus, we identify the quality of the metallic transport bridge as a crucial ingredient affecting device performance.
 
Figure \ref{fig:3}c presents a set of $VI$ curves for the devices NbTiN100 and NbTiN200. The colored solid lines are results obtained at 300 mK, while the gray horizontal lines highlight the boundaries of the C-band. We can observe that, although NbTiN100 has a higher $T_c^{wl}$, one has to carefully consider the operation conditions of the source, as the linear part of the resistive state of the $VI$ lies outside of the C-band at low-temperatures. The same is true for NbTiN200, however, as we demonstrate by the dashed line of matching color, it is not only possible to tune the $VI$ characteristics by magnetic fields, but also by temperature, gaining access to a more linear emission regime at 2 K for this device. 
Finally, Fig. \ref{fig:3}d confirms voltage-tunable radiation emission for NbTiN200 at 2 K, while we again observe a modulation of the measured power resulting from the transfer function of the normal transport bridge as discussed above. The differences in superconducting materials and fabrication processes confirm the robustness of the proposed approach of using JJAs as GHz radiation sources and allow us to envisage a better integration with future devices. Another important consideration is that at increasing temperatures, Johnson noise becomes more prominent while, at the same time, the power of the radiation decreases with decreasing $I_c^{wl}$  (see Supplemental Information). Consequently, it remains important to boost the power efficiency of the source and search for approaches that allow for coherent emission from the array.

\begin{figure}[h]
    \centering
    \includegraphics[width=1\linewidth]{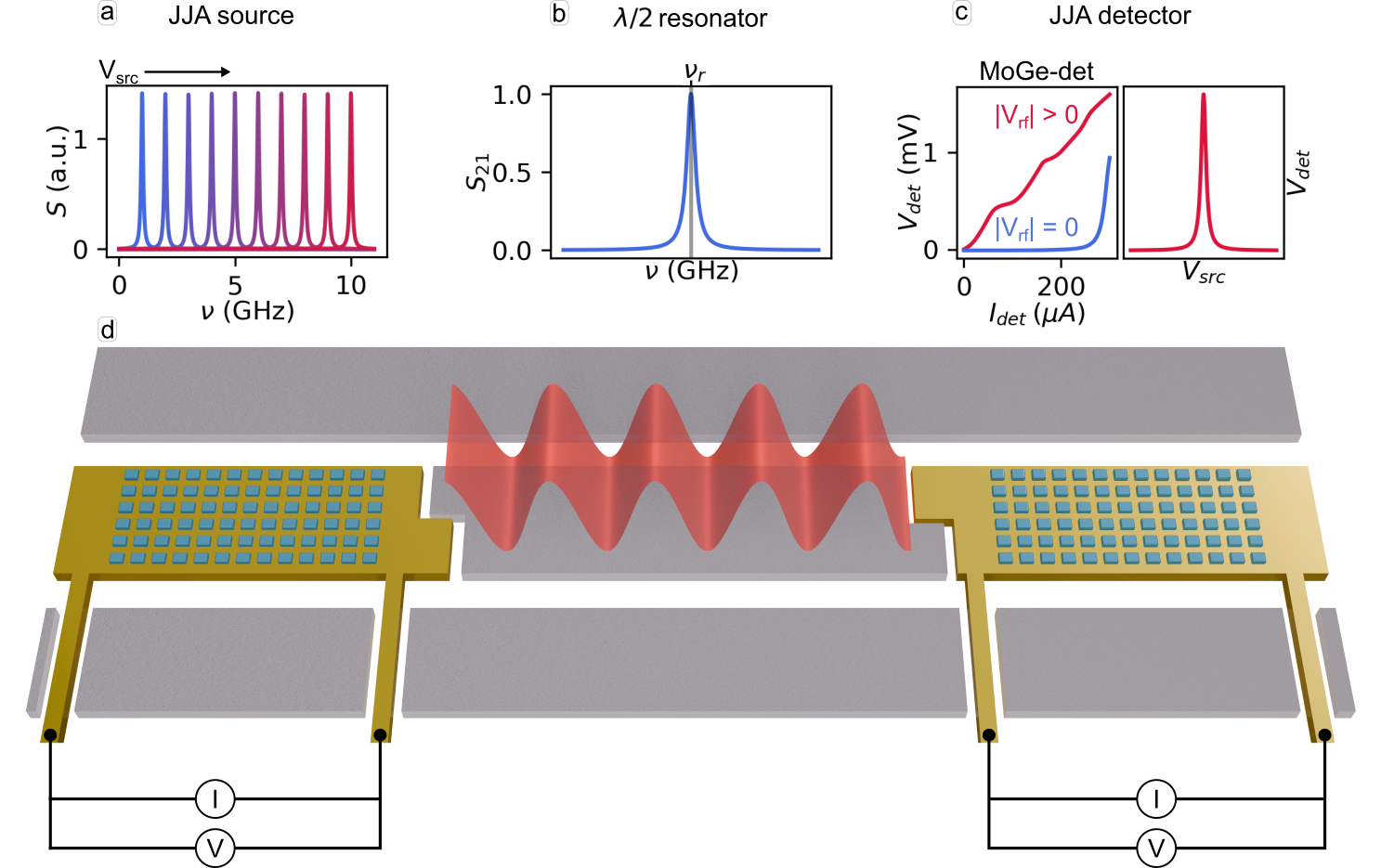}
    \caption{\textbf{Proposed on-chip microwave measurement platform using JJA as RF emitter and detector.} The top panels show \textbf{a} the ideal power spectral density of the JJA source at different voltages; \textbf{b} the transmission as a function of frequency for a $\lambda/2$ co-planar waveguide (CPW) resonator; and \textbf{c} the experimentally obtained $VI$ curves of the MoGe-det sample with and without RF irradiation together with the possible JJA detector response in the proposed platform. \textbf{d} A schematic representation of the DC-operated microwave co-planar waveguide platform consisting of a JJA source, a $\lambda/2$ resonator, and a JJA detector.}
    \label{fig:4}
\end{figure}

\section{Discussion}\label{Sc:Discussion}

In the previous section we fabricated JJAs that operate as GHz range radiation sources whose output frequency can be tuned through DC voltage, temperature and applied field. For all devices, we observe that poor coupling between the emitted signal and the output line compromises the ability to deliver such radiation to different points of a hypothetical circuit as would be necessary, for instance, to excite or readout a qubit. A substantial margin of performance improvement could be obtained by achieving coherent radiation emission, for which the observed linewidth should decrease below that of a single junction. A common strategy to improve the power efficiency of high frequency generators is to couple the emitting source to a cavity \cite{Cla73,Boo94,Bar99, Cas17}. However, the trade-off in this alternative design is the lack of the frequency tunability offered by the JJAs, as only resonant modes will be transmitted. Therefore, a more appealing approach is to optimize the design of the transport bridge to allow for better coupling with the radiation. A critical step towards this goal is to embed the arrays into an impedance-matched coplanar waveguide, avoiding losses to the environment and guaranteeing that the transmitted power is not modulated in frequency. This can be achieved with a similar material arrangement as the one presented here, with a normal metal transport bridge decorated with superconducting islands, as Au waveguides can transmit the signal with low losses \cite{Jud17}. An even better transmission could be achieved employing superconducting waveguides. In such case, the junctions need to be defined by etching the weak links in the superconducting film. In any case, numerical results based on finite element analysis can help to understand the ideal JJA-waveguide configuration maximizing the transmission and power efficiency of the devices. To achieve coherent emission and reduce the signal linewidth, it is crucial to consider the normal metal coherence length, a length scale describing the coherent transport of Cooper pairs through the junction. 
As Jain \textit{et al.} propose, this can be achieved by fabricating junctions with sizes in the order of the normal metal coherence length \cite{Jai84}.
An improved source that takes these issues into consideration is schematically represented in Fig. \ref{fig:4}a, highlighting the uniform and sharp radiation emitted when the source is biased at different voltages, $V_{\text{src}}$, akin to Fig. \ref{fig:2}b.

Yet, there is another enticing potential use of JJAs in the context of cryogenic high-frequency experiments. The emergence of giant Shapiro steps when the arrays are excited by an alternating signal allows these devices to be used also as a GHz range detector \cite{Cla73,Ben90,Hal90}. Figure \ref{fig:4}c demonstrates such features, as measured in the $VI$ of MoGe-det at 450 mK and $\tilde{f} = 0$ when it is irradiated by a 2 GHz signal with amplitude $\mathopen|V_{rf}\mathclose| = 1410$ mV. Meanwhile, at zero power, no Shapiro steps are visible if a DC current is driven through the array. When MoGe-det is irradiated by the RF signal, $I_c^{wl}$ is reduced from 210 $\mu$A to zero and giant Shapiro steps with a height of 4.14 $\mu$V reveal that all junctions in the array are synchronized by the external excitation. Therefore, by applying a bias measuring current $I_{\text{det}}$ below $I_c^{wl}$ and continuously probing the DC voltage $V_{\text{det}}$, it is possible to know if the detector array is being irradiated by an alternating signal or not \cite{Gri68,Tuc85}.

This possibility leads us to propose a novel GHz-range measurement platform. In such a device, two JJAs are embedded in a single transmission line: one acting as an ideal microwave source and the other as a microwave power detector. This is represented schematically in Fig. \ref{fig:4}d. As discussed throughout this text, both source and detector can be operated solely by DC circuitry. The waveguide transmits the signal from the ideal source to detector while allowing for the coupling of different physical systems to the radiation. For instance, let us consider the example of a $\lambda$/2 coplanar waveguide resonator that is capacitively coupled to the central waveguide, as shown in Fig. \ref{fig:4}. Such a component attenuates the transmitted radiation, except at a characteristic resonance frequency, $\nu_r$ \cite{Gop08}. Figure \ref{fig:4}b shows the typical $S_{21}$ transmission coefficient measurement of a $\lambda$/2 resonator, revealing a sharp increase in transmittance at $\nu_r$. The presence of the resonator then drastically modifies the signal arriving at the detector, suppressing the power if $\nu \neq \nu_r$, such that $V_{\text{det}}(V_{\text{src}})$ tends to zero at all frequencies different from $\nu_r$, as represented in Fig. \ref{fig:4}c. In this example, the detector is biased slightly below the critical current, where the sensitivity to changes in the $VI$ curve is larger. That way, the proposed device is able to probe the resonance frequency without requiring an external GHz source, a spectrum analyzer, a vector network analyzer, or any specialized radio frequency wiring or circuitry, drastically simplifying the requirements for spectroscopic measurements. In essence, the proposed measurement platform can be operated using only DC components and will allow GHz transmission measurements at cryogenic temperatures, replacing commonly employed RF components. Moreover, the frequency range is not limited to the C-band, but can be further extended up to the superconducting gap frequency, such that the device can be used, for instance, as a ferromagnetic resonance or electron spin resonance probe \cite{Mak15,Pla12}.

In summary, we have fabricated low-temperature GHz radiation sources that are operated solely by standard DC components. The sources are comprised of a SNS Josephson junction array which is coupled to a transmission normal metal bridge. We demonstrate that the emitted frequency is effectively tuned by the junction voltage according to the Josephson equations. The choice to employ JJAs in the device design allows us to further tune the source behavior by changing temperature, applied magnetic fields, the array arrangement, and material choice. We investigate the effect of all these parameters on the radiation while revealing that the coupling of the signal with the underlying transport bridge plays a determinant role in how the signal travels to different parts of the circuit. Moreover, we also demonstrate that the same arrays can be used to detect GHz range radiation, and propose a novel GHz range spectroscopy device fully operated by DC components. Such a device can substitute currently employed bulky and expensive RF circuitry by bringing source and detector together on the same chip, empowering future improvements in high-frequency characterization and quantum technology.

\section{Methods}\label{Sc:Methods}


\subsection{Sample fabrication}

The studied MoGe devices consist of Mo$_{78}$Ge$_{22}$/Au Josephson junction arrays fabricated on top of Ti/Au transport bridges grown on natively oxidized silicon (300 nm silica) substrates. The substrates are first spin-coated with a co-PMMA/PMMA positive resist double layer. The transport bridge patterns are transferred to the resist by electron beam lithography (EBL) and later developed for 35 s in a 1:1 MIBK:IPA solution. The reaction is stopped by introducing the sample in IPA for 30 s. Then, a 5 nm Ti sticking layer and a 30 nm Au layer are grown over the pattern by molecular beam epitaxy (MBE), both with a growth rate of 0.18 \AA/s. The transport bridge is revealed after lift-off in acetone. A second spin-coating and EBL procedure is conducted to define the array patterns. Following development, a 25 nm layer of Mo$_{78}$Ge$_{22}$ and a 5 nm Au layer are deposited by pulsed laser deposition with pulse width 5-8 ns at a repetition rate of 80 Hz. 
The final device is ready after a second lift-off procedure in acetone. The results presented in this paper are obtained from devices fabricated on three separate Si/SiO$_2$ substrates. Overall, we have verified radiation in the 4-8 GHz range arising from 7 devices fabricated in 5 separate substrates, confirming the reproducibility of the effect.
The NbTiN samples are fabricated on a 100-Si/SiO$_2$ wafer with a 400 nm oxide layer grown via a wet oxidation process. First, a 5 nm Ti adhesion layer is evaporated using electron beam evaporation in a physical vapour deposition system (Lesker PVD 225) followed by an in-situ deposition of a 25 nm Au layer.  Subsequently, a 25 nm NbTiN layer is sputtered on the gold surface using a near-ultra high vacuum magnetron sputtering system (DCA Metal Sputter System). The NbTiN pattern is defined using electron beam lithography in a Raith e-beam system with a negative-tone e-beam resist (ma-N 2403) which is developed in MF-319 developer.
The NbTiN is then etched using reactive ion etching in an Oxford Plasmalab 100 system with a CF$_4$/O$_2$ plasma mixture (90 sccm CF$_4$, 22.5 SCCM O$_2$) to achieve the desired pattern. Afterwards, the resulting structure is cleaned in MREM 400. The Au transport bridges are patterned using a subsequent round of EBL employing the same type of resist and developer as before. This pattern is etched in an Oxford Ionfab 300 system via secondary ion mass spectrometry, with a slight overetch to ensure complete removal of Au. Stable Si atom counts confirm this etching process. The resist is removed with MREM 400 and the final structures are verified with SEM imaging.

\subsection{Low-temperature transport and RF measurements}

\begin{figure}[h]
    \centering
    \includegraphics[width=0.6\linewidth]{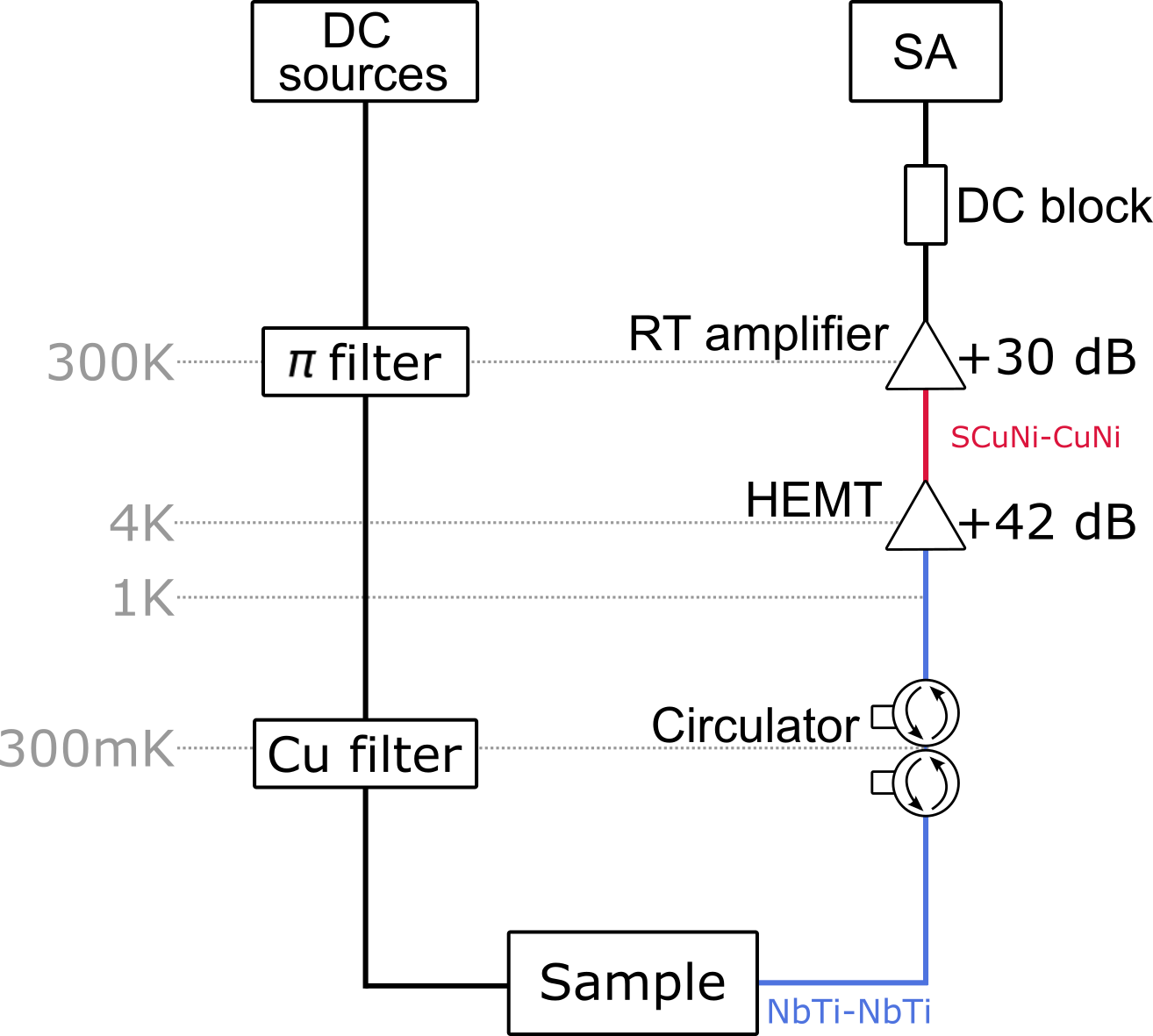}
    \caption{\textbf{Schematic representation of the cryostat used in the RF experiments} }
    \label{fig:meas}
\end{figure}

Low-temperature (0.3-4.2 K) measurements were carried out in a Janis $^3$He cryostat customized for measurements in the 4-8 GHz range, depicted in Fig. \ref{fig:meas}. Perpendicular magnetic fields are applied in the sample space using a superconducting coil. The input RF line is attenuated by 60 dB. The output line has a non-reciprocal double stage isolator (LNF-CICIC4\_8A 4-8 GHz Cryogenic Dual Junction Circulator) at the 0.3 K temperature stage, blocking external radiation. A high electron mobility transistor (LNF-LNC4\_8C) amplifies the signal with 42 dB at the 4 K stage. Both RF lines have a 50 $\Omega$ impedance. Additionally, eight DC lines allow to operate the devices and measure their $VI$ characteristics. A room-temperature $\pi$-filter with cutoff frequency of 1 MHz and a low-temperature LC copper powder filter are used to strongly attenuate unwanted signals through the DC lines in the 10 kHz to 10 GHz range. Bias currents are applied using a Keithley 2400 sourcemeter, while the sample voltage is monitored by a Keithley 2182A nanovoltmeter. The RF signal generated by the sample is amplified by 30 dB at room temperature and detected with a frequency sweep using a FSIQ26 signal analyzer from Rhode \& Schwarz.

\subsection{List of devices}
An overview of the samples with their relevant design parameters is shown in Table \ref{tab:devices}. The list contains the device number, the superconducting material, the position on the device (top or bottom) and geometrical parameters $d_x$, $d_y$, $S_x$, $S_y$ (see Fig. \ref{fig:1}\b) and $N_x$, while $N_y=14$. With the exception of Device 2, all devices have two JJA, one at the top and another at the bottom bridge, as represented schematically in Fig. \ref{fig:1}a.
\begin{table}[h]
\caption{\label{tab:devices} List of design parameters for all the reported samples, $N_y = 14$ is fixed for all devices.}
\begin{tabular}{c|cccccccc}
Device             & Material & Position & $d_x$ (nm) & $d_y$ (nm)& $S_x$ (nm)& $S_y$ (nm)& $N_x$ & Name          \\ \hline \hline
\multirow{2}{*}{1} & \multirow{2}{*}{MoGe}  & Top      & 300  & 200  & 500  & 500  & 51   & MoGe-top    \\
                          &                        & Bottom   & 300  & 200  & 500  & 500  & 51   & MoGe-bot \\ \hline
2                  & MoGe                   & -        & 270  & 125  & 510  & 570  & 125  & MoGe-det       \\ \hline
\multirow{2}{*}{3} & \multirow{2}{*}{NbTiN} & Top      & 200  & 200  & 500  & 500  & 51   & NbTiN200      \\
                          &                        & Bottom   & 300  & 200  & 500  & 500  & 51   & NbTiN300      \\ \hline
\multirow{2}{*}{4} & \multirow{2}{*}{NbTiN} & Top      & 100  & 200  & 500  & 500  & 51   & NbTiN100      \\
                          &                        & Bottom   & 400  & 200  & 500  & 500  & 51   & NbTiN400 \\  \hline\hline
\end{tabular}
\end{table}

\subsection{Data analysis}
Following the measurement of each emission spectrum, a background signal was recorded with a signal analyzer when no bias was applied, which is subtracted from the data. 
The resulting spectrum was then compensated to account for 72 dB of amplification in the output line.
The emission spectra were analyzed using Voigt function fits. While a single junction is expected to exhibit a Lorentzian lineshape, small Gaussian deviations in properties across the JJA can result in a broadning of the lineshape. Consequently, the Voigt function, which represents a convolution of a Lorentzian and a Gaussian, provided a more accurate representation of the observed lineshape. 

\subsection{White noise generation}
If the JJAs are biased by DC currents above $I_c$ of the MoGe islands, in the range of 6 to 6.5 mA and at 300 mK, no Josephson radiation is expected as all components of the device are in the normal state. Therefore, the dissipating currents running through the device will cause Joule heating in the metallic transport bridge, inducing Johnson noise, with a power spectral density $S = 4k_BTR$, independent of $\nu$ \cite{Dic46}. Through this process, the transport bridge acts as a source of white noise radiation, equally emitting at all frequencies.

\section*{Author Contributions}
S.V., D.A.D.C.  designed the devices. S.V., D.A.D.C., S. R., I.P.C.C., fabricated the devices. S.V., D.A.D.C. and M. A. executed the RF measurements presented in this manuscript. H.D., L.N. and B.R. designed the measurement setup, fabricated the prototypical devices and performed their full RF characterization. S.V, M.A. performed the simulations.  The manuscript was written by S.V, D.A.D.C.  and J.V.V. with the help from all other authors. All authors discussed the results and reviewed the manuscript. J.V.V., B.R., M.J.V.B and A.V.S.  initiated and supervised the research.

\section*{Acknowledgments}
This work is supported by Research Foundation Flanders (FWO) grant number 11K6525N and 11A3V25N, the EUCOST action SUPERQUMAP CA21144, the Fonds de la Recherche Scientifique - FNRS under the grant Weave -PDRT.0208.23 and CDR J.0199.25. This research is supported and funded by an interuniversity BOF project (IBOF-23-065). NbTiN samples were purchased from and fabricated by ConScience AB. The authors would like to thank Clécio C. de Souza Silva for the support with the RSJ model simulations.

\newpage
\section{Supplementary Information}
\subsection{Supplementary note 1: Impact of non-linear $VI$ characteristics on radio-frequency emission}
\begin{figure}[b]
\centering
\includegraphics[width=0.75\linewidth]{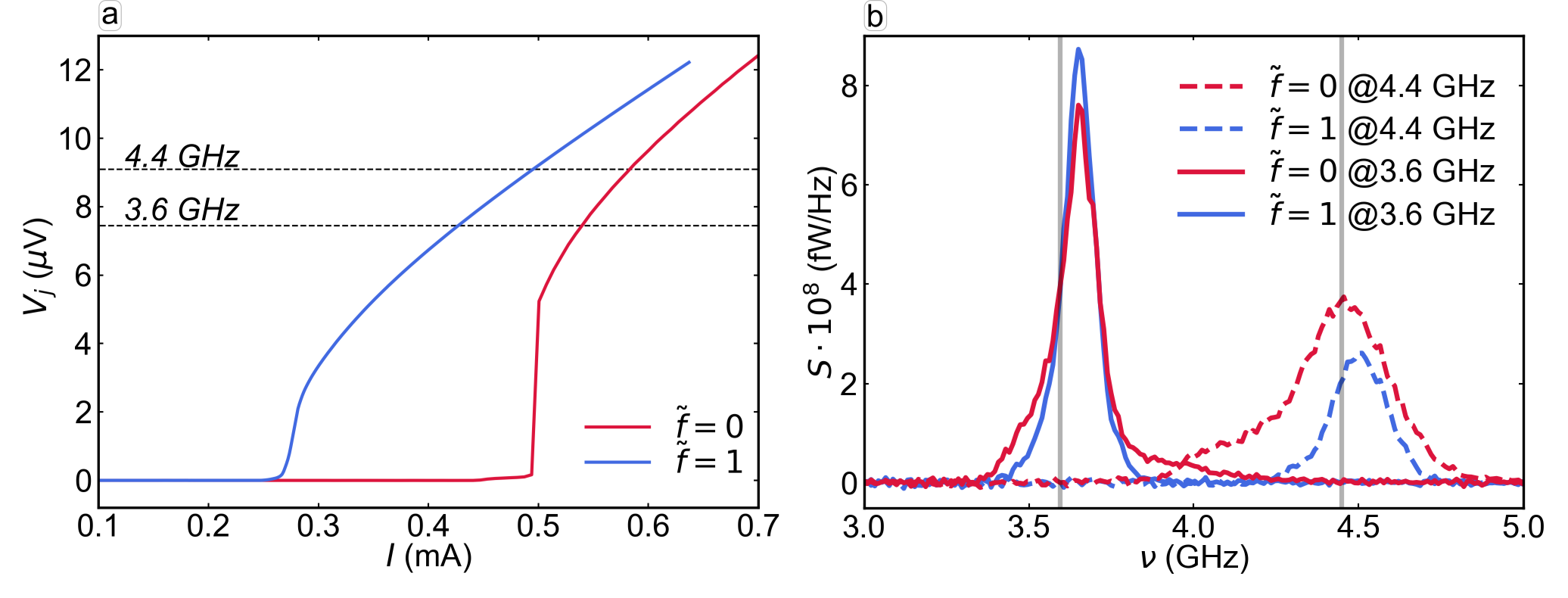}
\caption{\label{fig:SI1}\textbf{Effect of non-linear $VI$ characteristics on the emitted spectra. a} $VI$ curves at $\tilde{f}=0$ and $\tilde{f}=1$. \textbf{b} Measured spectra at 3.6 GHz and 4.4 GHz at $\tilde{f}=0$ and $\tilde{f}=1$. The grey lines represent the Josephson frequency $V_j/Phi_0$ for each peak.}
\end{figure}

In this section, we discuss the effect of the non-linear regime in the $VI$ curves on the measured spectrum. Apart from the occurrence of expected higher harmonics, we also notice that the FWHM increases in the non-linear regime of the $VI$-curve. The radiation is highly sensitive to voltage differences between the junctions, which is most pronounced in the non-linear regime. This becomes clear in Fig. \ref{fig:SI1}a which shows the $VI$ characteristics of MoGe-bot at frustrations $\Tilde{f}=0$ and 1. The horizontal lines indicate voltages corresponding to 3.6 GHz and 4.4 GHz. At these voltages, spectra are measured and shown in Fig. \ref{fig:SI1}b. The dashed lines are spectra measured at $\Tilde{f}=0$, i.e. when the $VI$ is non-linear. The full lines are spectra measured at $\Tilde{f}=1$, where the $VI$ is more linear. Comparing the spectra in these two situations, the emission is much broader in the non-linear regime at $\Tilde{f}=0$. Furthermore, at 3.6 GHz in the most non-linear part, the emitted power is reduced as well, compared to the spectrum in the more linear range of the $VI$ at $\tilde{f}=1$. Thus, the magnetic field can be used to tune the linearity of the $VI$-characteristics and consequently, the linewidth.

\subsection{Supplementary note 2: RSJ model}
The resistively-shunted junction (RSJ) model allows to accurately describe single SNS Josephson junctions \cite{Pan20,Rae20,Tin96}. The model is schematically shown in Fig. \ref{fig:SI2}a, where the junction can be modeled as a superconducting current channel governed by the Josephson equation
\begin{equation}
    I_s = I_c\sin{\phi},
\end{equation}
with $I_s$ the supercurrent, $I_c$ the critical current and $\phi$ the phase difference -- and a resistive quasiparticle channel with ohmic behavior
\begin{equation}
    I_N = V_j/R_N,
\end{equation}
with $I_N$ the quasiparticle current, $V_j$ the voltage over the junction and $R_N$ the junction resistance, which can be approximated as the constant normal state resistance for SNS junctions. The total current $I$ can be calculated using Kirchoff's current law
\begin{equation}\label{eq:SI_1}
    I = I_c\sin{\phi} + \frac{V_j}{R_N}.
\end{equation}
Equation \ref{eq:SI_1} can be rewritten in terms of the phase difference using the second Josephson equation ($V_j = \frac{\Phi_0}{2\pi}\frac{d\phi}{dt}$), leading to
\begin{equation}
    I = I_c\sin{\phi} + \frac{1}{R_N}\frac{\Phi_0}{2\pi}\frac{d\phi}{dt},
\end{equation}
which can be rewritten as
\begin{equation} \label{eq:SI2}
    \frac{d\phi}{dt} = \frac{2eI_cR_N}{\hbar}(i - \sin{\phi}),
\end{equation}
with $i = I/I_c$ the reduced current.
The time-dependent voltage $V_j(t)$ can be found by integrating equation \ref{eq:SI2} and using the second Josephson equation. The characteristic time of the junction is defined as:
\begin{equation}
    \tau_c = \frac{\hbar}{2eR_NI_c}.
\end{equation}When a DC current bias is applied with $i<1$, the current is purely carried by a supercurrent and the voltage is zero. As the bias current is increased to $i>1$, the current can no longer flow as a pure supercurrent and some of it has to flow through the resistive channel. This results in an oscillating voltage as a function of time, of which simulations are shown in the inset of Fig. \ref{fig:SI2}a. For low bias currents the $V_j(t)$ dependence is very anharmonic, while for higher bias currents it behaves almost like a perfect sine. The average voltage as function of applied current bias has the following relationship: $<V_j(t)> = I_cR_N\sqrt{i^2-1}$, shown in Fig. \ref{fig:SI2}a.

\subsection{Supplementary note 3: Theoretical emission power as function of applied voltage}
\begin{figure}
\centering
\includegraphics[width=1\linewidth]{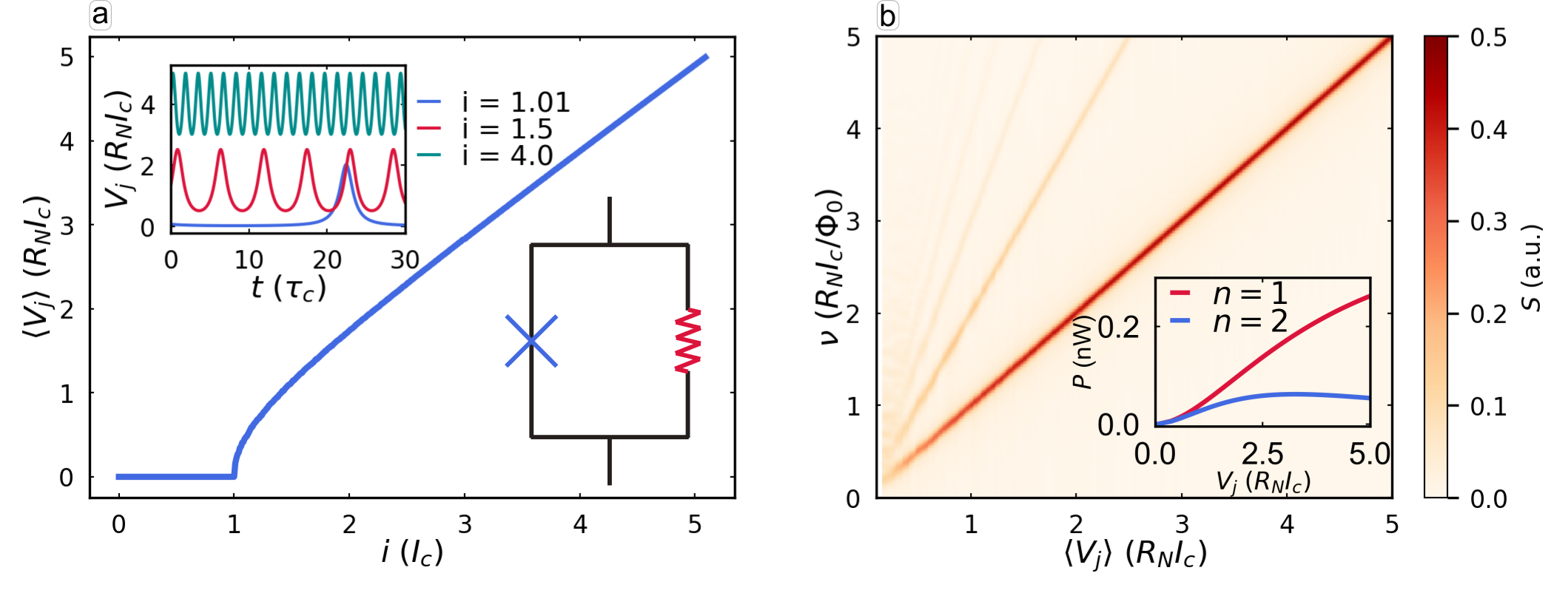}
\caption{\label{fig:SI2}\textbf{Results of the RSJ simulations. a} Simulated $VI$ curve using the RSJ model. The lower inset shows the circuit representation of the RSJ model. The upper inset shows the simulated time dependence of the oscillating voltage. \textbf{b} Radiation map from simulated RSJ model. The inset shows the power determined by the voltage oscillations $|V_n|$ of the voltage oscillations for the first ($n=1$) and second ($n=2$) harmonic.}
\end{figure}

Using the RSJ model, it is possible to simulate an emission map, shown in Fig.\ref{fig:SI2}b, from which the expected monotonic behavior of the emission power can already be seen.
When biased with a DC current, the oscillating voltage contains harmonics, of which the amplitudes are given by \cite{Jai84,Luk89}
\begin{equation}
\label{eq:eq1}
    |V_n| = V_c \frac{2\bar{v}}{(\sqrt{1+\bar{v}^2} + \bar{v})^n},
\end{equation}
where $\bar{v}=\sqrt{i^2-1}$, with $i=I/I_c$ the reduced current, $V_c = I_cR_j$ is the critical voltage, and $n$ the order of the harmonic. This relation is plotted in the inset of in Fig. \ref{fig:SI2}b for the first and second harmonic as a function of the applied junction voltage in units of GHz, where experimentally determined values of $R_j = 0.33$ $\Omega$ and $I_c = 495/14$ $\mu$A $= 35$ $\mu$A were used, resulting in a characteristic voltage of $V_c = 12$ $\mu$V. 
The amplitude of the first harmonic of the voltage oscillation increases monotonically, while that of the second harmonic decreases after a voltage corresponding to 2 GHz. As the emission power is determined by the amplitude of the voltage oscillations, the available power should have monotonical behavior as a function of applied voltage, which is in contrast to our experimental findings. 

\subsection{Supplementary note 4: Background signal as function of temperature}
\begin{figure}
\centering
\includegraphics[width=0.75\linewidth]{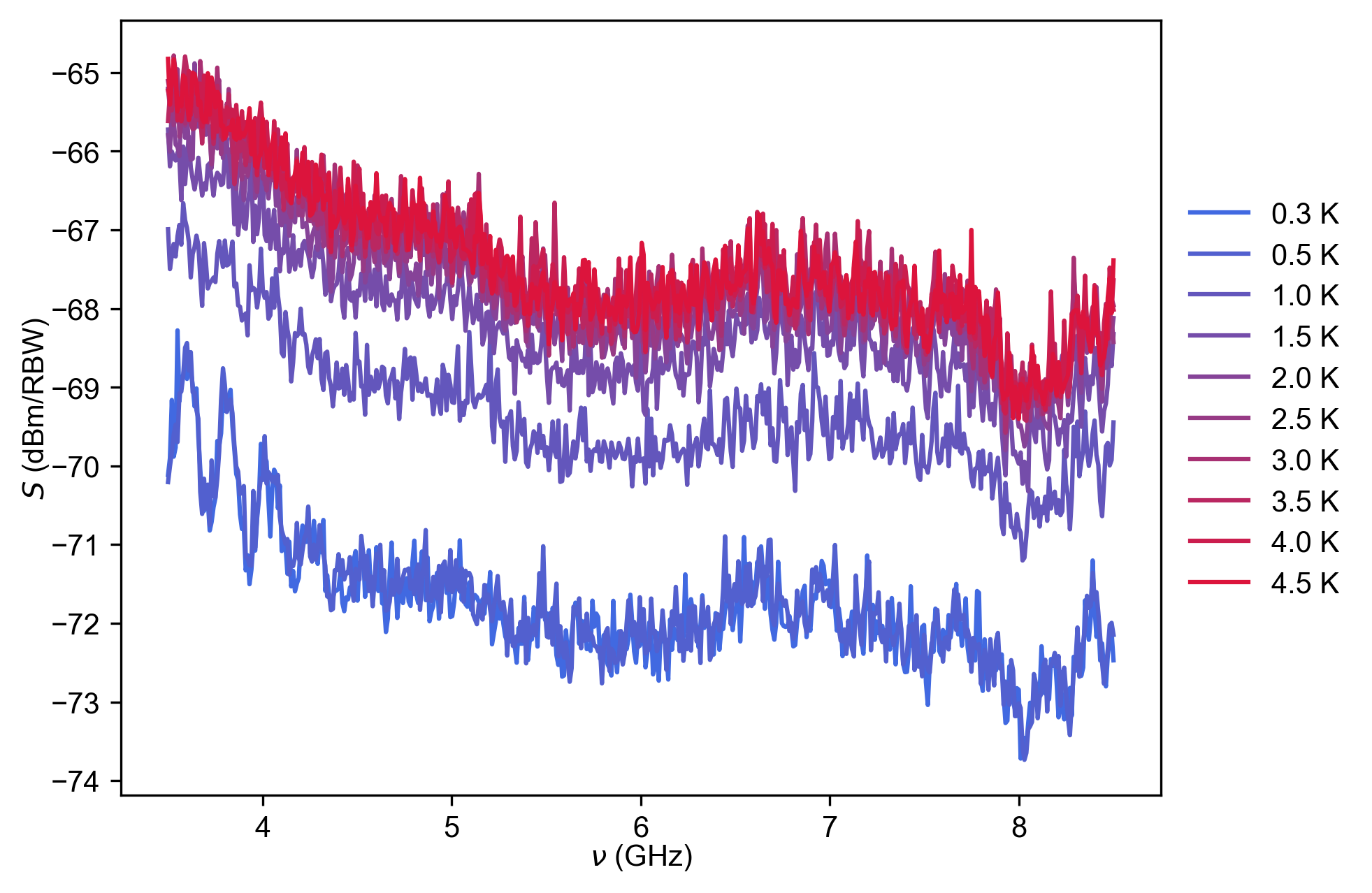}
\caption{\label{fig:SI3}Thermal background signal as function of frequency for temperatures between 300 mK and 4 K.}
\end{figure}

The thermal noise increases with temperature as can be seen in Fig. \ref{fig:SI3}, which shows the measured background power density signal as a function of frequency for different temperatures when no bias is applied to MoGe-bot. The combination of increasing background power with increasing temperature and decreasing power of Josephson emission with decreasing critical current [according to Equation (\ref{eq:eq1})] makes it difficult to measure radiation at 4.2 K.

\subsection{Supplementary note 5: SEM image of MoGe-top and MoGe-bot}
\begin{figure}
\centering
\includegraphics[width=0.75\linewidth]{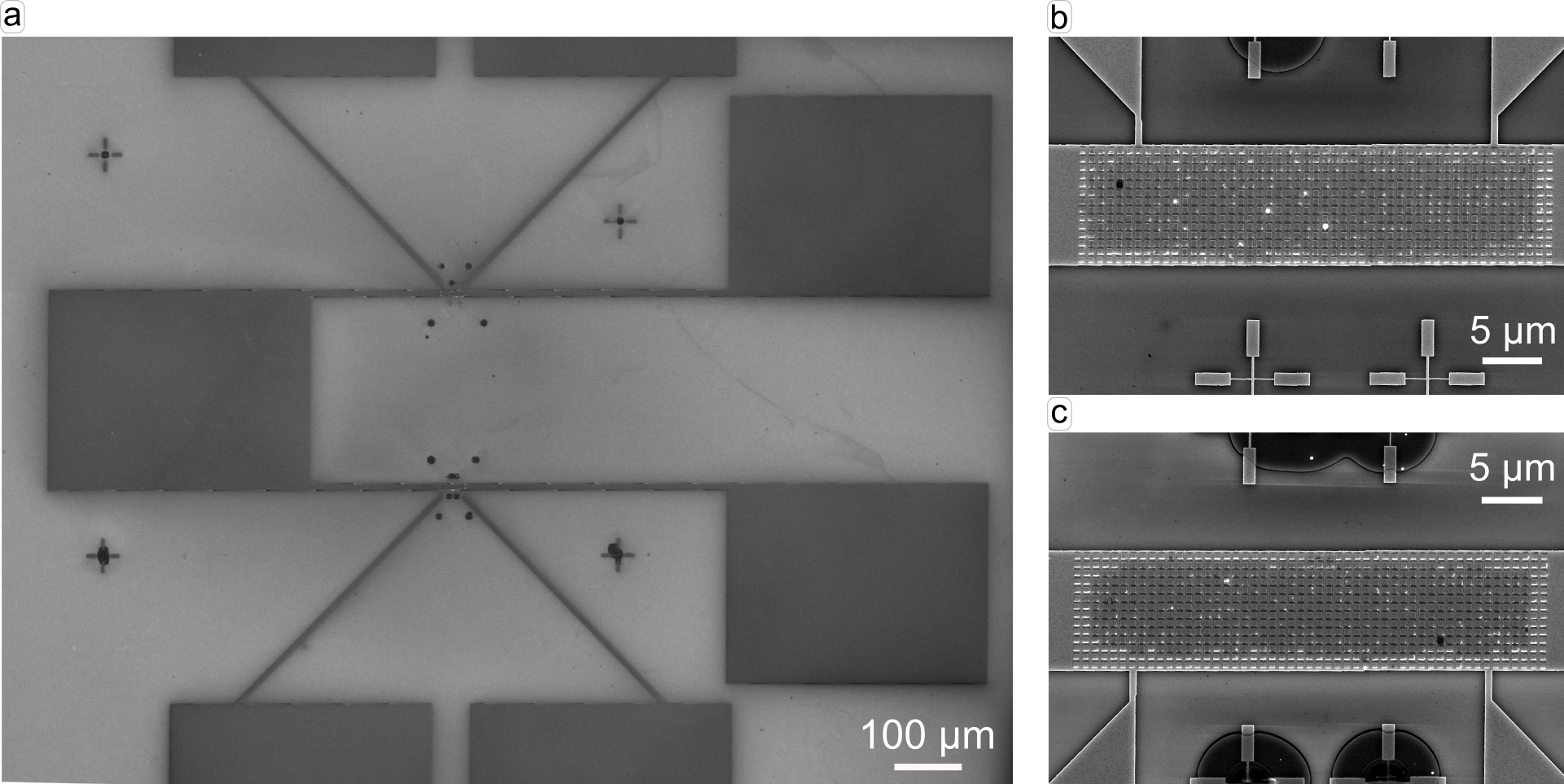}
\caption{\label{fig:SI4}\textbf{a} SEM image of Device 1. \textbf{b} and \textbf{c} are details of the JJ arrays of samples MoGe-top and MoGe-bot respectively.}
\end{figure}

Fig. \ref{fig:SI4} shows a SEM image of MoGe-top and MoGe-bot. Fig. \ref{fig:SI4}a shows the full structure, with both the top and bottom transport bridges in a single device. Fig. \ref{fig:SI4}b and c show details of MoGe-top and MoGe-bot, respectively.

\newpage

\end{document}